\documentstyle[epsfig]{article}
\everymath{\rm }
\everydisplay{\rm }

\begin{document}
\begin{center}
{\LARGE Long range correlations  \\ [.125in]
in quantum gravity} \\ [.25in]
\large Donald E. Neville \footnote{\large Electronic address:
nev@vm.temple.edu }\\Department of Physics \\Temple University
\\Philadelphia 19122, Pa. \\ [.25in]
March 17, 1998 \\ [.25in]
\end{center}
\newcommand{\E}[2]{\mbox{$\tilde{{\rm E}} ^{#1}_{#2}$}}
\newcommand{\A}[2]{\mbox{${\rm A}^{#1}_{#2}$}}
\newcommand{\Np}{\mbox{${\rm N}'$}}
\newcommand{\Etwo}{\mbox{$^{(2)}\!\tilde{\rm E} $}\ }
\newcommand{\Etld }{\mbox{$\tilde{\rm E}  $}\ }
\def \ut#1{\rlap{\lower1ex\hbox{$\sim$}}#1{}}
\newcommand{\phst}{\mbox{$\phi\!*$}}
\newcommand{\psist}{\mbox{$\psi\!*$}}
\newcommand{\bea}{\begin{eqnarray}}
\newcommand{\eea}{\end{eqnarray}}
\newcommand{\be}{\begin{equation}}
\newcommand{\ee}{\end{equation}}
\newcommand{\nn}{\nonumber \\}
\newcommand{\rta}{\mbox{$\rightarrow$}}
\newcommand{\rla}{\mbox{$\leftrightarrow$}}
\newcommand{\eq}[1]{eq.~(\ref{eq:#1})}
\newcommand{\Eq}[1]{Eq.~(\ref{eq:#1})}
\newcommand{\eqs}[2]{eqs.~(\ref{eq:#1}) and (\ref{eq:#2})}
\large
\begin{center}
{\bf Abstract}
\end{center}
Smolin has pointed out that the spin network formulation of quantum
gravity will not necessarily possess the long range
correlations needed for a proper classical limit; typically, the
action of the scalar constraint is too local. Thiemann's length
operator is used to argue for a further restriction on the action
of the scalar constraint: it should not
introduce new edges of color unity into a spin network, but
should rather change preexisting edges by $\pm$ one unit of
color.  Smolin has proposed a specific ansatz for a correlated
scalar constraint.  This ansatz does not introduce color unity
edges, but the [scalar, scalar] commutator is shown to be
anomalous.  In general, it will be hard to avoid anomalies, once
correlation is introduced into the constraint;  but it is
argued that the scalar constraint may not need to be anomaly-free
when
acting on the kinematic basis.  \\[.125in]
PACS categories: 04.60, 04.30
\clearpage

\section{Introduction}

     The recently proposed spin network formulation of quantum
gravity \cite{symmsta, ALMMT, Baez} is very appealing, because it
is similar to
Wilson loop approaches \cite{GambTri} which have been used very
succesfully in
quantum chromodynamics.  Wilson loops are inherently non-local, and
even though it is possible to construct non-local objects within
the traditional approach utilizing quantized local fields, the spin
network approach has focused attention on the non-local structures
which seem to be needed to satisfy the constraints of quantum
gravity.  

     Although local fields are used in the initial construction of
spin network states and operators, the final result does not
involve local fields.  The physical meaning of a given state is
encoded in the SU(2) ``colors''
assigned to a network of edges (one-dimensional curves embedded in
three-dimensional space); and in the way in which these edges are
connected together at vertices where edges meet.  The
operators of the theory alter the edges and
vertices in specified ways, without explicitly invoking local
fields.  

     In a Dirac constrained quantization framework, the spatial
diffeomorphism constraints are
relatively easy to satisfy, using spin networks, because
information encoded in the edge colors and vertex connectivities is
not sensitive to the way the edge curves are coordinatized in the
underlying three-dimensional spatial manifold.  Also, research into
spin networks has led to greatly increased understanding of how to
regulate quantum operators \cite{AV, Lollvol, diffreg}. With
hindsight, this understanding could have been
achieved within the framework of standard quantum field theories.
(The key is to consider only operators which respect spatial
diffeomorphism invariance.  In the language of differential
geometry, the key is to smear n-forms over n-surfaces
\cite{volAL}.)  However, this insight into
regularization was not achieved in the traditional framework, and
the new approach should be given the credit for  stimulating a
great deal
of new thinking.

     There are disadvantages to an approach based on spin networks,
rather than local fields.  Smolin \cite{corr} has pointed out that
any approach
which downplays the
continuum and relies on discrete structures (spin
networks; Regge calculus) will not automatically posess the long
range correlated behavior  needed for a satisfactory classical
limit. In the Regge calculus approach, long-range behavior emerges
from discrete dynamics only when Newton's constant and the
cosmological constant are tuned to critical values such that a 
correlation length diverges in the limit of small edge lengths
\cite{Regge}.  

     In the spin network approach, one does not approach the
continuum by varying a length.  One recovers the classical limit,
and the classical
fields, by demanding that edge
lengths are small compared to the length scale over which the
underlying fields vary appreciably. 

     In this limit, consider the scalar constraint, which is the
hardest constraint to satisfy and the
hardest constraint to understand intuitively.  The classical limit
does not determine the spin network scalar constraint uniquely:
the literature contains two 
different spin network operators \cite{Thiesc, SmoRov, Rov} which
reduce to the same Dirac scalar
constraint in the
limit of small edge lengths.  In this situation, one falls back on
simplicity as a criterion to
choose among the possibilities; but, as
emphasized by Smolin,  the spin network scalar constraints
proposed up to now are too local in character to give rise to long
range correlations.  More precisely, the scalar constraint produces
both local and non-local effects, but the latter do not give rise
to long range correlations.  Local effects: when  the scalar
constraint 
acts at a vertex, the constraint typically changes the color of
each edge as well as the
connectivity of the vertex, that is, the way in which the SU(2)
quantum numbers of the edges are
coupled together to form an SU(2) scalar.  This action is
completely local (confined to a single
point, the vertex).  Non-local effects: if
the edges radiating from a given spin network vertex are visualized
as a set of spokes radiating
from a hub, then the constraint adds an edge of color unity which
connects a pair of edges
meeting at the
vertex, so that (if the constraint acts several times) the wheel
begins to look like a spider's web.  Thiemann calls these
connecting
edges of color unity ``extraordinary edges'' \cite{Thiesc}.   This
action is non-local because the extraordinary
edges are attached to the spokes at a finite distance from the
vertex.  The non-local action
produces no correlation, because, if one moves out along an edge
from the original vertex to a
neighboring vertex, the colors and connectivity at the neighboring
vertex have not changed: the
neighboring vertex has no way of ``knowing''  that anything has
happened at the original vertex. 
This situation is illustrated in figures~1 and 2, which show a
portion of a spin
network before and after the scalar constraint acts at the upper
vertex. \\

\begin{figure}[bht]
\centerline{\mbox{\epsfig{file=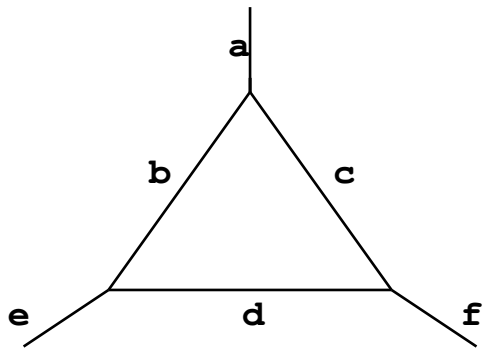,bbllx=0in,bblly=0
in,bburx=3in,bbury=2in}}}
\caption{A spin network, before action of the scalar constraint}
\label{fig1}
\end{figure}
\begin{figure}[bht]
\centerline{\mbox{\epsfig{file=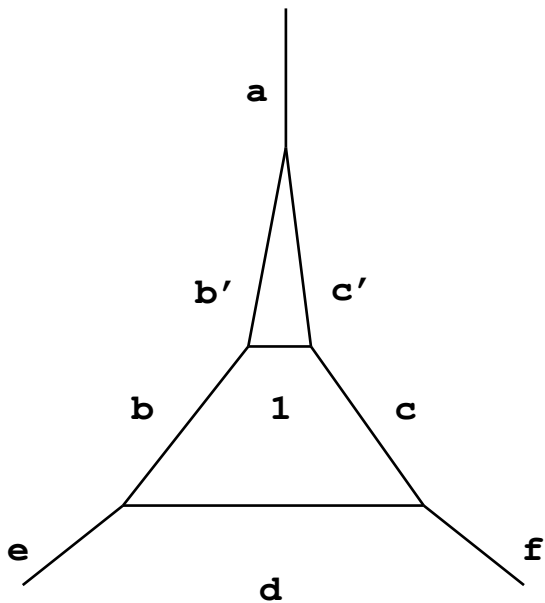,bbllx=0in,bblly=0
in,bburx=3in,bbury=3.5in}}}
\caption{Spin network of fig. 1, after scalar constraint has acted
at upper vertex}
\label{fig2}
\end{figure}
(The spin network shown in the figures is a bit special, in that
all vertices are trivalent, with only
three edges meeting at each vertex; this does not affect the
argument.)  The scalar constraint has
changed the colors at the vertex where it has acted (local action),
and has altered the diagram at
some distance from the vertex by adding an extraordinary edge
(non-local action); but if one
moves out along the left-hand edge starting from the top vertex and
ending at the bottom
left-hand vertex, the latter vertex has not been changed.  The
action is non-local, but
nevertheless uncorrelated.

     Assuming the scalar
constraint is Hermitean, there
will also be matrix elements which {\it remove} one extraordinary
edge; and
a state in the kernel of the
constraint presumably would be an infinite series of states having
ever increasing numbers of
such edges.  However, no matter how many extraordinary edges the
state
contains, if A and B are two
nearest neighbor vertices
(connected by an edge) then  whether or not the constraint is
satisfied at vertex A depends
entirely on how the edges are arranged at A; the arrangement at
vertex B is irrelevant.   This does not resemble classical gravity,
where what is going on at the
sun is supposed to affect what is happening at the earth.  
Presumably in order to introduce  correlations one must  make the
scalar constraint less local in character. 

     The present paper has four parts. The first part, at the end
of  this introduction, gives another
argument, based on
Thiemann's recently introduced length operator \cite{Thieln}, that
the (non-local, but uncorrelated)  spin
network scalar constraints
constructed up to now are not physically plausible because they
dramatically distort length
relationships within the spin network.  The second part, in
section II of this
paper,  explores what seems to be the simplest way of introducing
correlation 
without distorting
lengths: expand a certain color unity loop occuring in the
definition of the scalar constraint, so
that the loop fills an entire triangle of the spin network, not
just a portion of the triangle.   

     The third part of the paper, in
section III,  considers a specific  recipe for a correlated scalar
constraint, one essentially identical
to the recipe proposed by Smolin, and shows that  this specific
choice has a 
scalar-scalar commutator which is anomalous.  The last part of the
paper, in section IV, discusses attempts to determine the scalar
constraint by constructing a four-dimensional spin network
formalism.   

     Note that the usual, uncorrelated choices for the
scalar constraint are anomaly-free, almost trivially so. 
The usual choices for the scalar constraint C act on one vertex at
a time, so that C may be
written as a linear sum of operators, one for each vertex A, B,
$\cdots$ in the spin network.
\be
     C = C_A + C_B + \cdots.
\label{eq:1.a}
\ee
In the scalar-scalar commutator C occurs smeared by arbitrary
scalar functions M
and N.
\bea
     [\int MC, \int NC]& =& [M_A C_A + M_B C_B + \cdots ,
                    N_A C_A + N_B C_B + \cdots] \nn
                    &=& (M_A N_B - M_B N_A)[C_A,C_B] + \cdots .
\label{eq:1.b}
\eea
$ M_A$ is the value of the smearing function at vertex A, etc.  The
scalar-scalar commutator
becomes a series of cross terms $[C_A,C_B]$ which vanish because
the action of C at vertices A
and B is not correlated.  (The scalar-scalar commutator  must
vanish, not just equal the spatial
diffeomorphism constraint, because the commutator  is acting on the
spin network basis, which
is in the kernel of the diffeomorphism constraint.)  Once the
constraint is made correlated, the vanishing of
$[C_A,C_B]$ is no longer automatic.  
Requiring the constraint operator to be anomaly-free does not
restrict an uncorrelated constraint, but
should strongly determine any correlated constraint.

     Although I spend some time in section III to show that the
specific recipe considered
there is anomalous, it is not altogether clear to me that freedom
from anomalies is a reasonable
criterion to impose.  In the classical theory, the commutator
algebra must be anomaly-free, in
order for the theory to possess full diffeomorphism invariance
\cite{HKTei}.  Intuitively, the various
three-dimensional time slices can be ``stacked'  so as to form a
manifold with four-dimensional,
not
just three-dimensional  diffeomorphism invariance.   It might seem
natural to demand an
anomaly-free commutator in the spin network case; but the spin
network states form a basis
which is only kinematical, not physical.  (Each member of a
kinematical basis is annihilated by
the Gauss SU(2) internal rotation constraints and the spatial
diffeomorphism constraints, but is
not necessarily annihilated by the scalar constraint.  The physical
basis is that subset of the
kinematical basis which is also annihilated by the scalar
constraint.)    Since the kinematical
states are not physical, in general, there are no  observational
consequences even if the
commutator is anomalous.  (For example, in a path integral, the
unphysical states are excluded
from the path integration by appropriate functional delta
functions, so that the unphysical states
do not even occur as virtual states.)  Further, on the physical
subset, where there might be
observational consequences, the commutator [C,C] is trivially
non-anomalous, since each factor
of C separately annihilates the state.  Additionally, the spin
network representation in effect
replaces the continuum with a one-dimensional subset of edges and
vertices.  Perhaps one should
not be surprised if full diffeomorphism invariance becomes
difficult to implement in this
situation.  

     I give up the requirement of anomaly-free commutators with
some reluctance, because
the specific recipe for introducing correlations discussed here is
not unique.  (One alternative
possiblilty  is discussed  in section II.)  It is desirable to
impose  freedom from anomalies, in
order to determine the scalar constraint as fully as possible.  If
this requirement is not imposed,
then at present it is not clear what physical requirement  fully
determines the scalar constraint.

     Currently Thiemann's expression for the scalar constraint
seems to be  the one most
widely accepted \cite{Thiesc}.  The Thiemann form makes extensive
use of the volume operator, because that
operator is invariant under spatial diffeomorphisms and therefore
is relatively easy to regulate. 
However, in other respects the Thiemann constraint is very hard to
work with.  It is the sum of
an Euclidean constraint, which involves one volume operator, plus
a ``kinetic''  term, which
involves three volume operators; and the volume operator itself is
quite complex.   Accordingly, 
in section II, when demonstrating the presence of anomalies, I will
not try immediately to
generalize
Thiemann's constraint to a more correlated version.   Instead, I
will work with a generalization
of the Rovelli-Smolin constraint \cite{SmoRov, Rov}, essentially
the generalization suggested by Smolin \cite{corr}.  This
means I will have to pretend that the analytic factor in the matrix
element can be regularized in a 
diffeomorphism invariant way, whereas in fact no one knows how to
regulate this factor. 
(By ``analytic factor'  I mean every factor in  the matrix element
except the 
group theoretic factor contributed by the SU(2) dependence of the
vertex.)    Nevertheless, the
calculation with the Rovelli-Smolin form should not be misleading,
since the Rovelli-Smolin
and Thiemann forms share certain crucial elements in common (in
particular, the presence of
color unity loops in the expression for the constraint operator). 
Once the Rovelli-Smolin
form has been discussed in detail, in section II, it is
straightforward to show how to introduce
correlations into the Thiemann form of the
constraint.  The discussion of anomalies in section III also uses
the Rovelli-Smolin form, but
again  this should not be misleading, because the anomaly is
generated by the group theoretic
factor, not by the lack of diffeomorphism invariance in the
regulator.  

     I now discuss Thiemann's length operator briefly, then use it 
to argue that any scalar constraint operator which introduces color
unity edges will signifigantly distort length relationships within
the spin network.  The Thiemann length operator is one of
three operators recently proposed to
measure geometrical properties (length, area, and volume) of a spin
network \cite{
AV,volAL,Thieln,area,vol}  .  All three operators
are spatially diffeomorphic invariant, so that any non-invariant
structures introduced to regulate
these  operators drop out of final results.  I will assume that all
three operators are consistent
with Euclidean geometry, and with each other.  For example, suppose
one measures the volume
of a spin network tetrahedron using the volume operator, then
measures the length of each side
of the tetrahedron using the length operator.  From Euclidean
geometry, the volume is also given
by  a  determinant constructed from the lengths.  For consistency,
the volume measured directly
should equal the  volume computed from the lengths.  There is no
particular reason to expect
consistency, except  for  spin networks which approximate classical
states.   (Eventually, one
may even be able to determine classical states by demanding that
the geometric operators are
consistent, when acting on these states.)  For the argument which
follows, I will assume that all
lines in the spin network, except the extraordinary lines added by
the scalar constraint,  have 
color much greater than unity, so that the state can be assumed to
be classical, the geometric
operators are consistent, and it is reasonable to use intuition
based on Euclidean geometry.

     Further, I will assume all vertices are trivalent (three edges
meet at each vertex); or at
least, the network contains a subset which is entirely trivalent,
because the spectrum of the
length operator has been computed for such vertices.  The
restriction to trivalent vertices is
probably not essential.  Figure~1 shows such a trivalent
subset consisting of six edges. The labels a through f are the
colors of the edges; all colors
are assumed to be order n, $n \gg 1$.  
Figure~2 shows the same
subset after the scalar constraint has acted once at the upper
vertex and inserted an extraordinary edge.  In figure~2 and
succeeding figures, the notation $b'$ stands for b~$\pm$~1;
similarly $c'$ = c~$\pm$~1, etc.; the action of the scalar
constraint changes figure~1 into a weighted sum over the various
possibilities for $b'$ and $c'$.  In figure~2 I have suppressed a
summation over $b'$ and $c'$ as well as the weighting coefficients.

     I have drawn the upper half of the triangle as squeezed to a
much smaller area, because
the length operator predicts this is what happens to the triangle:
the color unity line is short, and is inserted near
the midpoints of edges b and c, rather than near the
vertex.  The lines labeled
b and $b'$ = b~$ \pm~1$, for example, both have lengths of order 
$l_p$ n, $l_p$ the Planck length;
while  the color unity line has length of order $l_p \sqrt{n}$,
very short compared to all the other
lines in the diagram.  This is intuitively a very implausible
result.  In the classical limit one thinks of the scalar constraint
as almost commuting with other operators, such as the length
operator.  This implies the scalar operator should produce only a
very small fluctuation in the geometry of the state, typically
changing lengths by order $l_p$, and therefore  areas by  $\Delta
A$/A = order 1/n. 

     I now verify in detail the statements made above about the
lengths of edges in figure~2.  If an edge joins two
trivalent vertices, then Thiemann has shown that the squared length
of the edge is a sum of two contributions, one from each vertex. 
For example, for edge b in figure~2,
\be
          L^2(b) = \lambda^2(b; e, d) 
                    + \lambda^2(b;b \pm 1, 1),
\label{eq:1.2}
\ee
where
\bea
   & & (4/l_p^2) \lambda^2(c;a,b)  = \nn
& & \; \frac{2c+ 1}{2c+1/2} [(a+b+c+1)
                         (a+b-c) (-a+b+c+1/2) (a-b+c+1/2)]^{1/2}
\nn
& & \; +\frac{2c}{2c+1/2} [(a+b+c+1/2)
                         (a+b-c+1/2) (-a+b+c) (a-b+c)]^{1/2} .
\label{eq:1.3}
\eea  
The formula \eq{1.3} is very complicated in general, but in the
limit that a, b, and c are large,
\be
     \lambda^2(c:a,b) \rta l_p^2 A(a,b,c),
\label{eq:1.4}
\ee
where A is the area of the triangle with sides (a/2,~b/2,~c/2).  In
the limit
that one of the edges a, b, or c is an extraordinary edge, \eq{1.4}
is not quantitatively accurate, but does give the correct order of
magnitude of   $\lambda^2$.  Thus the contribution to $L^2$ from a
vertex with all
three edges of order $n \gg 1$ is order $l_p^2 n^2$; while the
contribution to $L^2$ from a vertex with one extraordinary edge is
order $l_p^2 n$.  In figure~2  $L^2(b)$ is the sum of
a contribution of order $l_p^2 n^2$ from the vertex at the lower
end of b, plus a negligible
contribution
of order $l_p^2 n$ from the vertex with the extraordinary edge at
the upper end of b. 
Similarly for the edge $b'$ = b~$\pm$~1; therefore the two edges
b and $b'$  in figure~2 both have lengths
of order $l_p n$, and the vertex with the extraordinary edge occurs
at the approximate midpoint of the side.  The extraordinary edge
itself is much shorter, since both vertices contribute order $l_p^2
n$ to $L^2$, threrefore order $l_p \sqrt{n}$ to L.  This completes
the demonstration that lenght relationships are strongly distorted
when a color unity line is added to a spin network.  

\section{Correlations}

     From the discussion in the Introduction, any scalar constraint
recipe which introduces
color unity edges into a diagram will distort lengths.  Also, if
the color of an edge is not
modified  along its entire length, the constraint will affect only
the vertex at one end of the edge,
and there will be no correlation.  The obvious solution to both
these problems is to push the
color unity edge in figure~2 all the way to the bottom of the
triangle, until it coincides
with edge d; then recouple so that the color one edge disappears,
and the bottom edge
has color $d~\pm~1$.  This is the qualitative idea, expressed in
graphic terms.

     To become more quantitative, one must revert to the underlying
local field
theory, construct the operators and the states in terms of  local
fields, then infer the
corresponding spin network operators and states, for both the
correlated and uncorrelated
version of the scalar constraint operator.   The earlier steps in
this procedure have been discussed
at length
in the literature \cite{symmsta,ALMMT,volAL, looprep} , and I will
repeat those discussions only enough to recall  a few key steps
in the procedure.  Also, I will not try to rederive the SU(2)
recoupling factors which occur at
some steps.  

     There is universal agreement as to the
field theoretic meaning of the spin network state: a holonomy
matrix h is associated with each
edge of color c in a spin network.
\be
     h(\gamma) = Pexp[i\int ds \dot{\gamma}_a (s) A^A_a S_A],
\label{eq:2.1}
\ee
where P denotes path ordering, $\dot{\gamma}_a $ is the tangent
vector to the edge, the
integration is over the entire edge, the $A^A_a$ are the
connections for the rotational (SU(2))
subgroup of the local Lorentz group, and the $S_A$ are  the
generators of  SU(2) for the
irreducible representation having color c.  An explicit factor i is
needed because I take the generators
$S_A$ to be  Hermitean.

    Associated with each vertex is an SU(2) 3J symbol which couples
the
(suppressed) $S_Z$ indices on the h's to a total spin zero.  (There
will be more than one 3J
symbol if the vertex is higher than trivalent.)  

     There is less agreement as to the field theoretic meaning of
the operators, since the
classical limit does not determine the operators uniquely.  I work
primarily with the
Rovelli-Smolin prescription for the scalar constraint, which is 
\be
     C_{RS} = -Tr [\E{[a}{A}(x_1) \sigma_A h_{ab}([x_1,x_2])
\E{b]}{B}(x_2) \sigma_B         
           h_{ab}([x_2,x_1])],
\label{eq:2.2}
\ee
where \E{a}{A} is the densitized inverse triad, the momentum
variable conjugate to \A{A}{a},
and the two $h_{ab}$ together form a small closed loop in the ab
plane.  Of course the
small loop serves to point split what would otherwise be a poorly
defined product of \Etld field
operators, and the holonomies h must be inserted to keep the
construction SU(2) invariant; but in
addition the loop holonomies supply a  factor of  $F_{ab}$ needed
for the classical limit, F the
field strength constructed from the connection \A{A}{a}.  If
$h_{ab}([x_1,x_2])$ is a very short
segment of  loop, so that
most of the area of the loop is enclosed by $h_{ab}([x_2,x_1])$,
then in the classical limit
where
the fields are varying slowly over the area of the loop, 
\bea
     h_{ab}([x_1,x_2]) &\approx& 1; \nn
     h_{ab}([x_2,x_1]) &\approx& 1 + iF^C_{ab} \sigma_C
(area)/2;\nn
     C_{RS}&\approx&\E{a}{A}  \E{b}{B} F^C_{ab} \epsilon_{ABC}
(area),
\label{eq:2.3}
\eea
where (area) is the total area enclosed by the loop.  Assuming all
the loops $h_{ab}$ are given
the same area, independent of ab, one can divide out
the area factor and arrive at (almost) the usual classical scalar
constraint.  ($C_{RS}$ has the
wrong density weight to yield a diffeomorphism invariant when
integrated over $d^3x$.  As
mentioned earlier, this non-invariance  leads to difficulties  when
the constraint is regulated.)

     Now continuing to work in the field theoretic language,  allow
$C_{RS}$, \eq{2.2}, to act on
the state.  Since the \Etld are canonically conjugate to the A
fields, \E{a}{A} acts on (or
``grasps'' ) each holonomy in the state  like a functional
derivative $\delta/\delta\A{A}{a}$, and
brings down a group-theoretic factor of $S_A$ from the exponential
of the holonomy.  The
$S_A$  is
multiplied by  $\sigma_A$  from the scalar constraint, times
analytic factors which I ignore.  
Because the tensors $S_A$ and $\sigma_A$ each carry two suppressed
matrix indices in
addition to  the color two index A, S and $\sigma$ are essentially 
3J symbols which couple two
color c
lines to form a color two line (in the case of S) or two color
unity lines to form a color two line
(in the case of $\sigma$).  

     Finally,  translate this action back into the language of spin
networks:   each grasp by an
\Etld introduces two vertices
into the spin network (the two 3J symbols corresponding to $S_A$
and $\sigma_A$), while the
$h_{ab}([x_2,x_1])$  in the scalar constraint introduces a small
loop of color unity.  Figure~3 shows the spin network which results
when $C_{RS}$  acts
at the upper vertex of figure~1.   
\begin{figure}[htb]
\centerline{\mbox{\epsfig{file=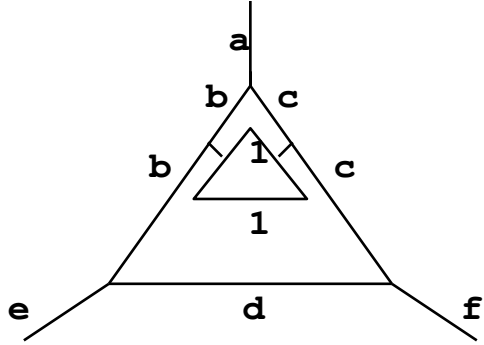,bbllx=0in,bblly=0.2
in,bburx=3in,bbury=2.2in}}}
\caption{Action of constraint on upper vertex of fig. 1}
\label{fig3}
\end{figure}
Note \Etld ($x_1$) and \Etld ($x_2$) must grasp two different edges
in figure~3, because
the antisymmetry of the scalar constraint in the indices ab kills
terms where the two \Etld grasp
the same edge.  Also $x_1$ and $x_2$ are close together, which
means the two grasps must
occur close to a vertex, as shown in the figure.  In fact all five
vertices at the top of figure occur
at exactly the same location.  They are drawn slightly separated
for clarity, but there is no
holonomy on any of the edges connecting any of the five vertices. 
  (The small color unity line
at the top of the diagram carries the holonomy $h_{ab}([x_1,x_2])$,
which is approximately
unity.)  SU(2) recoupling theory
may be used to rearrange the 3J symbols, therefore, until the
diagram resembles figure~2. 
(I am still working with an uncorrelated constraint and have not
yet constructed the correlated
constraint.)  To obtain figure~2, one  recouples the four colors
connected to the color two
line on the left  (colors 1, 1, b, b), as well as the four colors
connected to the color two line on
the right (colors 1, 1, c, c); the result is  figure~4. 
\begin{figure}[htb]
\centerline{\mbox{\epsfig{file=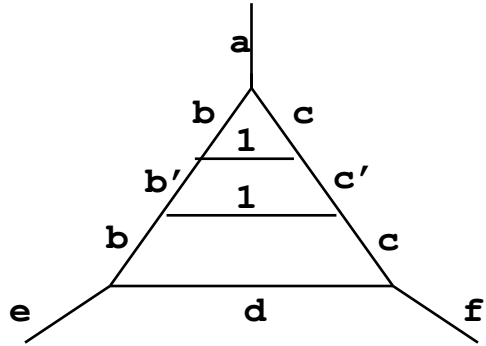,bbllx=0in,bblly=0.1
in,bburx=3in,bbury=2.1in}}}
\caption{Fig. 3 after two recouplings}
\label{fig4}
\end{figure}
The spin network of figure~4 should be multiplied by two 6J symbols
from the recoupling
of the b and c lines; I suppress these for the moment.
The five vertices at the top of figure~4 are still at the same
point, but one can in effect
shift the two lowest of these vertices downward in space by
shifting an holonomy onto the $b'$ and $c'$
lines: factor the two holonomies on  the color unity and color b
lines and
slide one factor from each line upward to form an holonomy on the
color $b'$ line; do the
same for the color $c'$ line.   In this way one shifts the two
lowest vertices downward to
the midpoints of the b and c lines.  If now one recouples to remove
the small triangle with sides
b,c,1,   the result, the uncorrelated scalar constraint action, is
the spin network of
figure~2.

     It is  straightforward to modify the above procedure to
produce a correlated constraint: do
not factor the b and unity holonomies; slide both  holonomies
entirely upward onto the $b'$ edge, so that the lower (b,1,$b'$)
vertex moves all the way down the b side, to the lower
left vertex of the original triangle.  Perform a similar maneuver
on the c side.  The result is
shown in figure~5.
\begin{figure}[htb]
\centerline{\mbox{\epsfig{file=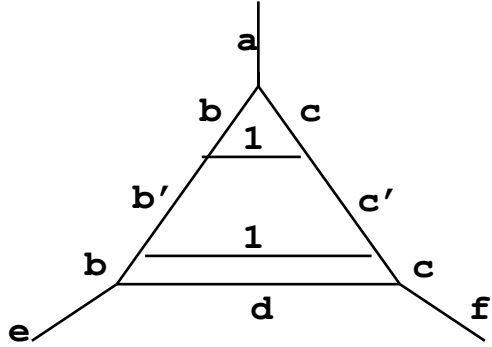,bbllx=0in,bblly=0
in,bburx=3in,bbury=2.1in}}}
\caption{Action of a correlated Rovelli-Smolin constraint on upper
vertex of fig. 1}
\label{fig5}
\end{figure}

     If one wishes, one can now remove the color unity line
entirely from the diagram. 
Recouple the color d and color unity edges at the b end (or at the
c end; it does not matter) as
shown in figure~6.
\begin{figure}[htb]
\centerline{\mbox{\epsfig{file=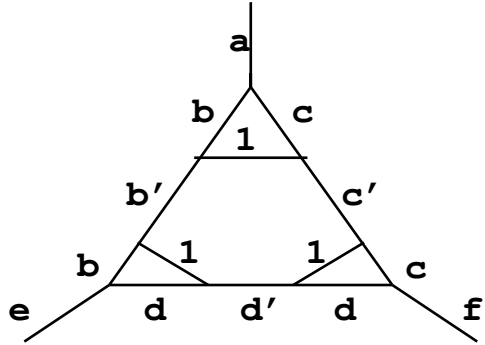,bbllx=0in,bblly=0.1
in,bburx=3in,bbury=2.1in}}}
\caption{A recoupling which removes the color unity line}
\label{fig6}
\end{figure}
Slide the holonomies from the color d and color unity lines onto
the color $d'$ = d~$\pm~1$ line; this in
effect moves the right-hand (d, 1, $d'$) vertex all the way to the
right and replaces the
original pair of holonomies by a single holonomy having color d
$\pm 1$.  There will still be
color unity lines in  two small triangles at each end of the new
color d~$\pm~1$ line; these
triangles can be removed by a further recoupling.  

     The correlated version of the constraint will have the same
classical limit as the
uncorrelated version, provided the second line of \eq{2.3}
continues to hold.  That is, the fields
must be slowly varying over the entire area of the triangle bcd,
not just over the upper half of the
triangle.  

     One must also ask about the (area) which formerly was merely 
an overall constant to 
be divided out of the scalar constraint, last line of  \eq{2.3}. 
This (area) equals the area
enclosed by the color unity loop, now  the entire area of the
triangle.  This area is no longer 
necessarily infinitesimal and can vary from triangle to triangle. 
 One can continue to divide it
out:  compute the  area of each triangle by using the Thiemann
length operator to compute the
length of the sides; then use standard trigonometry to compute the
area from the lengths;  then
(write the scalar constraint as a sum of terms, one for each area,
and) divide each term in the
scalar constraint by the  area.  The resultant formula  is very
ugly. I will argue that worrying
about this factor amounts to taking the Rovelli-Smolin example  too
seriously, since the
Thiemann recipe for the scalar constraint does not suffer from this
problem.   Moreover,  the
area factor probably should {\it not} be divided out, since the
scalar constraint always occurs
multiplied by $d^3x$.   Writing $d^3x$ = d(area) d(length), one
sees that a factor of area
belongs in the constraint. 
The real problem is not  how to remove the area, but  how to
include a length.  

     Clearly the Rovelli-Smolin constraint does not have a
wonderful analytic factor; but it does
have a group-theoretic factor which is relatively simple, yet
sufficiently complex to be
informative.  Both the Thiemann and the Rovelli-Smolin constraints
contain \Etld operators
which ``grasp'  the state, introducing vertices with color two
lines into the spin network;  and
both constraints contain an holonomy which introduces a color unity
loop into a triangle of the
spin network.  Following Smolin, I have introduced correlation by
expanding the color unity
loop to fill the entire triangle.  The same procedure works for the
Thiemann case.  The
Thiemann constraint is a sum of two Euclidean scalar constraint
operators, plus a kinetic term. 
The Euclidean operators are 
\bea
     C_{E1} &=& const. \epsilon^{abc}Tr\{(h_{bc}-h_{cb})
h_a[h_a^{-1},V]\};\nn
     C_{E2}&=& const. \epsilon^{abc} Tr\{
h_a[h_a^{-1},V](h_{bc}-h_{cb})\}.
\label{eq:2.4}
\eea
$h_{bc}$ and $h_{cb}$ are color unity loops of opposite orientation
lying in the bc plane. 
(Orientation counts since the indices on h are not traced over.) 
In the classical limit these loops
produce the factor of field strength F, hence are entirely
analogous to the color unity loop of the
Rovelli-Smolin example.  The third holonomy, $h_a$, is a straight
line segment parallel to
external edge a, hence  $h_a$  introduces no length distortions,
since it does not join two sides.  
Consequently, there is no obvious need to move it, when
constructing the correlated version of
the constraint, and I leave it alone.  (Its commutator with the
volume operator V is needed to
turn the three-grasp operator V into a
two-grasp operator, essentially the  \E{a}{A}  \E{b}{B} factor in
the classical limit, \eq{2.3}.)  

     I will not write down the kinetic operator, because it is very
complex; but the important
point is that there are no loops.  It involves three commutators of
the form [operator, $h_a$],
where again,  $h_a$ is a straight line segment parallel to an
external edge, hence  introduces no length distortions.   Even
though the kinetic term contains no
loops, it will be correlated.  Two of the three operators needed
for the kinetic term are formed by
taking commutators of the volume operator with $C_E$; hence $C_E$
induces correlated
behavior in  the kinetic term.

     I have also considered another procedure for introducing
correlation: rather than increase
the size of the loop, increase the distance between the grasps.  In
the $C_{RS}$  example, I took
the two grasps to occur at essentially the same point, which is why
one of the holonomies in
\eqs{2.2}{2.3} reduces to the unit matrix.  With a little more
work, one can move the grasps a
finite distance apart, away from a vertex to the middle of  edges,
so that neither holonomy is a
unit matrix, yet the constraint continues to have  the correct
classical limit.  Moving the grasps a
finite distance apart  tends to introduce lines of color unity into
the middle of the triangle,
however.  
Since these lines introduce unacceptable length distortions, one 
must move them (by reshaping
them until they are parallel to the sides, then recoupling, then
sliding holonomies, as before) 
until the color unity lines reach  edges or vertices of the
triangle and can be recoupled away.  By
this point the color unity lines are back at the vertices, and one
might as well have started out by
grasping at  vertices.  One has obtained nothing new.

     It is possible to get something genuinely new by changing the
recouplings which led from
figure~3 to figure~4: recouple the four colors (b,b,1,1) as shown
in figure~7;
and similarly,  recouple the four colors (c,c,1,1).
\begin{figure}[htb]
\centerline{\mbox{\epsfig{file=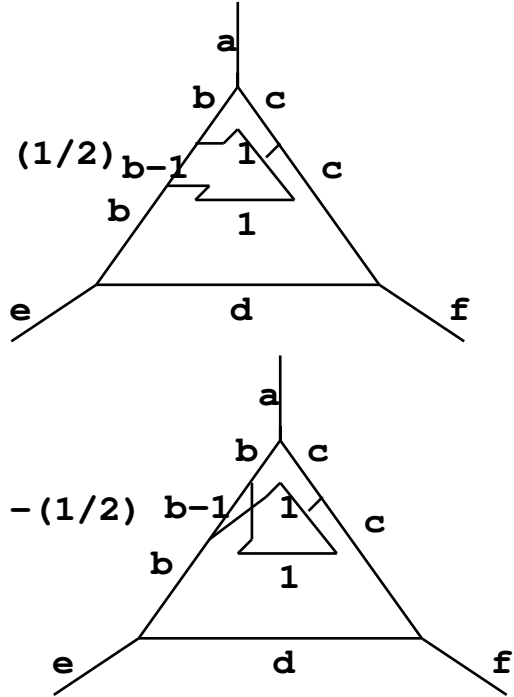,bbllx=0in,bblly=0
in,bburx=3in,bbury=3.8in}}}
\caption{An alternative recoupling of the b line}
\label{fig7}
\end{figure}
(The recoupling of figure~7 will be recognized as a standard way of
rewriting  a color two
line as a product of two color unity lines.)  The single spin
network diagram expands into four
diagrams.   All four have two parallel color unity lines crossing
the top of the triangle; when
these lines reach edge b or c, the two lines either cross or do not
cross.   One  parallel line comes
from the $h_{ab}([x_1,x_2])$ holonomy in \eqs{2.2}{2.3} (the upper,
slightly shorter line in
each of the four diagrams); the
other line comes from the $h_{ab}([x_2,x_1])$ holonomy (the lower,
longer line).   Now one
must get rid of these two color unity lines, first pulling them
from the middle of the diagram
until they are parallel to the edges.  If one always pulls the
lower line downward, upper line
upward, then 
this duplicates the motion which was used previously, and one
obtains nothing new (although
the final spin
networks will look superficially different because they have been
recoupled differently).  To
obtain
something new, move {\it vertices} rather than lines.  There are
two (b,$b-1$,1) vertices on the b
line, and similarly for the c line.   Move the two top vertices
upward, and the two bottom
vertices downward.  Some of the color unity lines now begin at the
top vertex and end at a 
bottom vertex.  It is easy to see that this gives something new,
because one can route
the lines from top to bottom along the sides in such a way that
there is no diagram with sides (b+1, c+1).  I will not pursue this
variant further, because the
lack of a final (b+1,c+1) state is, if
anything, a little too good to be true.  One expects the scalar
constraint to be as complex as
possible.  However, this variant  teaches an important lesson: the
prescription advocated here
(expand the spin unity line to fill the entire triangle) does not
lead to a unique final spin network.

\section{Anomalies}

     In order to check for anomalies, one needs the
group-theoretic factors which arise when
the scalar constraint acts on a spin network.  Fortunately, the
same factors occur repeatedly.
Consider the factors which arise when the scalar constraint acts
on the spin network of
figure~1 to produce figure~5.  These factors are
\be
     bc \left\{ \begin{array}{ccc}
                b & b & b \pm 1\\
                1 & 1 &     2    
                \end{array}
                              \right\} 
          \left\{ \begin{array}{ccc}
                    c & c & c \pm 1\\
                    1 & 1 &    2
                   \end{array}
                              \right\} 
          \left\{ \begin{array}{ccc}    
                    b       & b \pm1 & a \\
                    c \pm 1 &   c    & 1
                  \end{array}
                              \right\}                           
          \theta(b,c,a) / \Delta_a.
\label{eq:3.1}
\ee
The first factors, bc, I will call ``grasp  ' factors, because
they arise when the \Etld operators in
the scalar constraint initially grasp the b and c lines.  One can
think of the b line as made up of b
parallel color unity lines, and the factor of b arises from b
identical diagrams where the \Etld
grasps each color unity line in turn; similarly for the factor c. 
The next two curly brackets in
\eq{3.1} are the 6J symbols which arise when the four edges
(b,b,1,1) and (c,c,1,1) are
recoupled.   This is the recoupling which changes  figure~3 into
figure~4.   I will
call these 6J symbols ``color two recoupling factors'' , since
initially the b and l lines are 
connected by a color two line.  The product of the final 6J
symbol, the $\theta$ function, and the $\Delta_a$ function may be
denoted collectively the
``triangle''  factor, since these three factors arise when the
small triangle with sides (b,c,1) is
removed from the upper vertex in figure~4, to yield the simpler
upper vertex shown in
figure~2 or figure~5.  $\theta(b,c,a)$ is essentially the square of
a 3J symbol, while $\Delta_a = (-1)^a (a + 1)$.

     Only one other group theoretic factor is needed, a ``color
zero recoupling factor''  which
arises when at a later step the two lines at the bottom of
figure~5 are recoupled to give
figure~6:
\be
     \Delta_{d \pm 1}/ \theta(d,1,d \pm 1).
\label{eq:3.2}
\ee
This may be called ``color zero''  recoupling, because the initial
two parallel lines d and 1 are not
linked by any color.  

     General formulas for the 6J and $\theta$
functions have been worked out by
Kaufmann and Lins \cite{KL}, and specific numerical values have
been tabulated by De Pietri and Rovelli \cite{vol}.  Persons
trained in traditional recoupling theory may be slightly puzzled by
the 6J recoupling factors given in \eqs{3.1}{3.2}: the ``3J'' and
``6J'' symbols used here possess the same symmetries as the
traditional 3J and 6J symbols, but are normalized differently
\cite{KL,vol}. 

     For
the final spin network state of highest weight ($b~\pm~1 = b~+~1$,
and similarly all other
$\pm1 = +1$), the factors in \eqs{3.1}{3.2} are especially simple:
\bea
     \left\{ \begin{array}{ccc} 
               b& b& b + 1\\
               1& 1&    2
             \end{array}
                         \right\} &=& 1/2; \nn
     \left\{ \begin{array}{ccc}  
               b     &  b +1 & a \\ 
               c + 1 &   c   & 1
             \end{array}
                    \right\} \theta(b,c,a) / \Delta_a &=& 1; \nn
       \Delta_{d + 1}/ \theta(d,1,d + 1) &=& 1.
\label{eq:3.3}
\eea
The recoupling and triangle factors are just constants; only the
grasp factors are functions of the
color arguments.  This simplicity suggests the following strategy
for discovering anomalies. 
When the [scalar,scalar] [C,C] commutator acts upon a spin network
such as that
of figure~8, the result
is  a linear combination of spin networks with modified edges $b~
\pm~1$, $c~\pm~1, \cdots$; and
perhaps some edges with $d~\pm~2, \cdots$, since two scalar
constraints act on the network and
can increase or decrease the same color twice.    
\begin{figure}[htbp]
\centerline{\mbox{\epsfig{file=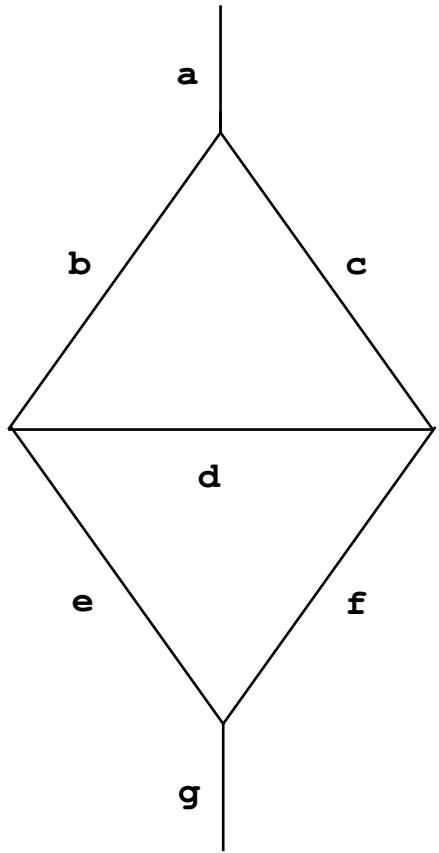,bbllx=0in,bblly=0.1
in,bburx=3in,bbury=4.8in}}}
\caption{A two-triangle subset of a spin network}
\label{fig8}
\end{figure}
Consider a term in the linear combination such as figure~9, in
which all colors have been
increased, never decreased.  This spin network will be multiplied
by the simplest group theoretic
factor, and whether or not the  factor  vanishes should be
relatively easy to determine.  
\begin{figure}[htbp]
\centerline{\mbox{\epsfig{file=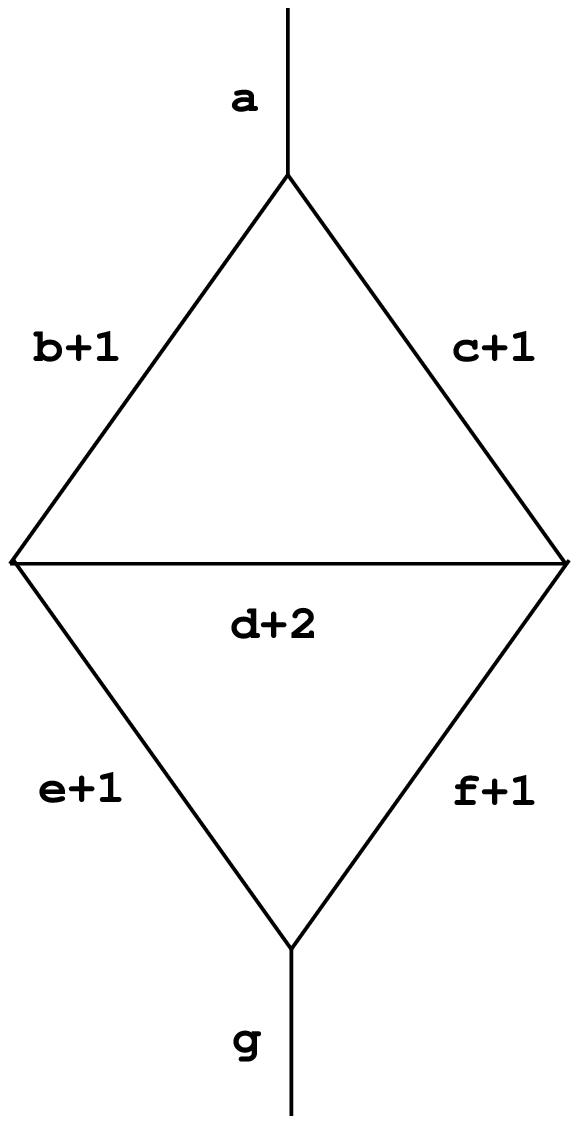,bbllx=0in,bblly=0
in,bburx=3in,bbury=4.8in}}}
\caption{A final state such that all colors have increased}
\label{fig9}
\end{figure}

     Having chosen a definite final state from all those occuring
in [C,C], the next step is to choose how the constraints
are  smeared (how the Lagrange multipliers are evaluated).  I will
consider two procedures: vertex smearing and area smearing.  Vertex
smearing is the procedure used at \eqs{1.a}{1.b}: multiply the
constraint
by the value of the smearing
function (Lagrange multiplier) at the vertex which is grasped.
In \eq{1.b}, for example, $M_A$ is the value of the
smearing function at vertex A, and $C_A$ is the sum of all terms
in the scalar constraint which
grasp some pair of edges ending at A.  Now that the small loop is
extended over an entire
triangle, however, perhaps a more appropriate procedure is to use
area smearing: 
multiply the constraint by the
average value of M on the triangle, for instance by $M_{bcd} =
(M_B + M_C + M_D)/3$, for a 
triangle with sides bcd and vertices BCD.  The commutator, 
\eq{1.b}, would be replaced by
\be
      [\int MC, \int NC] = (M_{bcd} N_{def} - M_{def}
N_{bcd})[C_{bcd},C_{def}] +
                         \cdots ,
\label{eq:3.4}
\ee
where $C_{bcd}$ is the sum of all terms in the scalar constraint
which grasp a pair of edges in
the triangle bcd.  Since the functions M and N are arbitrary, the
commutator must vanish term by
term.  Each $[C_A,C_B] = 0$ for vertex smearing; and each
$[C_{bcd},C_{def}] = 0$ for area
smearing.  

     Vertex smearing is easier to check, since
$C_A$ contains fewer terms than
$C_{bcd}$; therefore check vertex smearing first.  Let vertex A be
the
topmost vertex in
figure~7; let vertex B be the leftmost vertex, where edges bde
meet.  Since I want only
those grasps
which lead to the final state of figure~9, I need to consider
only the term in $C_A$ which
grasps sides bc, and only the term in $C_B$ which grasps sides
de.  The result is
\be
     [C_A,C_B](figure~8) = \cdots + (figure~9)[bcde -
(d+1)ebc](1/2)^4 .
\label{eq:3.5}
\ee
The $(1/2)^4$ comes from the four color two recoupling factors;
all color zero recoupling
factors and triangle factors are unity.  Only the grasp factors
(in the square bracket) are non-zero. 
 They do not 
cancel, and  the commutator is  anomalous.  It is easy to see why
this happens.   When the grasp
across lines de occurs first ($C_A C_B$ term in the commutator)
the grasp factor is de; whereas
when the grasp across lines de 
occurs second ($-C_B C_A$ term in the commutator) the grasp
factor is (d+1)e, because the
prior action of $C_A$ has changed the color of edge d to d+1.  
Obviously this grasp anomaly is
likely to occur whenever there is correlation: whenever the
action of C at one location (vertex or 
triangle) changes the colors of the lines at another location. 
For the  general final state
there will be recoupling and triangle anomalies as well as grasp
anomalies.  Once correlation is
introduced, anomalies are the rule, rather than the exception.  

     Now consider area smearing.   Consider the
$[C_{bcd},C_{def}] $ term.   $C_{bcd}$ is
a sum of three terms, since the scalar constraint can grasp edges
bc, cd, or db.  All three grasps
can change the b, c, and d  sides of figure~8 upward by one unit. 
Similarly, all three
grasps in  in $C_{def} $ can change the d, e, and f sides upward
by one unit; hence all six terms
making up  $C_{bcd}$ and $C_{def} $ will lead to the final state
of figure~9, and all
terms must be kept.  Fortunately, again, only grasp factors are
non-trivial.  The final result is 
\bea
  & &[C_{bcd},C_{def}](figure~8) = \cdots \nn
     & & \;+ (figure~9)\{[bc +b(d+1)+c(d+1)][ef + ed + fd] \nn
      & &\; - [ef + e(d+1) + f(d+1)][bc + bd + cd] \}(1/2)^4.
\label{eq:3.6}
\eea
 This expression is anomalous also.

     As discussed in the introduction, it is not entirely clear
that imposing freedom from
anomalies is a reasonable thing to do.   However, one might wish
to impose the requirement that
the anomalous terms in the commutator are small, in the classical
limit, because in the classical, 
continuum theory a very wide range of fields may be used to carry
representations of the
diffeomorphism group, including fields which do not satisfy the
scalar constraint.  Presumably
anomalous terms are ``small'  if  the commutator [C,C] has small
matrix elements compared to
matrix elements of $C^2$, in the limit that all colors a, b, c,
$\cdots$ are order n, $n \gg 1$.
From \eq{3.5} or \eq{3.6}, this requirement is satisfied, since
\be
     [C,C]/ C^2 = O(1/n).
\label{3.7}
\ee

\section{Four-dimensional approaches and further discussion}

     Reisenberger and Rovelli \cite{ReisRov} have suggested that a
kind of 3+1 dimensional ``crossing symmetry'' could be used to fix
some of the arbitrariness in the scalar constraint.  They construct
a proper time coordinate T and propagate a spin network from proper
time 0 to T using a path integral formalism.  Each path is weighted
by an exponential $exp[-i\int_0^T dT' H]$, where H is the usual
gravitational Hamiltonian, a sum of constraints .  They expand the
exponential in powers of H, and visualize the action of each power
of H on the spin network as follows.  (The diagrams are drawn in
2+1 rather than 3+1 dimensional spacetime for ease of
visualization, and the vertical direction is the proper time
direction T.)  Stack figure 2 vertically above figure 1.  Figure 1
represents a portion of the spin network at T = 0, before H has
acted; figure 2 represents the same spin network at proper time T
after H has acted.   Connect figures 1 and 2 by three approximately
vertical lines.  All three lines begin at the ``a'' vertex of
figure 1.  One line connects the ``a'' vertex of figure 1 to the
``a'' vertex of figure 2; the other two lines connect the ``a''
vertex of figure 1 to the two new vertices at the ends of the color
unity line in figure 2.  The action of the scalar constraint
therefore introduces a tetrahedron into spacetime: the bottom
vertex is the ``a'' vertex of figure 1; the three vertical edges
are the world-line of this vertex and the world lines of the two
new vertices introduced by H; the top face is the spin network
triangle with edges $(b',c',1)$ in figure 2.  For later reference
note that the three vertical edges of this tetrahedron are
qualitatively different from the three horizontal edges
$(b',c',1)$.  The vertical edges are world-lines of vertices; they
are not colored; they are not edges of any spin network.   

     Any time slice (horizontal slice) through the middle of the
tetrahedron will separate it into a ``past'', ccontaining one
vertex, and a ``future'', containing three vertices.  Reisenberger
and Rovelli call this a (1,3) transition.  They argue that 3+1
diffeomorphism invariance should allow one to rotate the
tetrahedron in 3+1 dimensional space, or equivalently time-slice
the tetrahedron at any angle.  This is what they mean by ``crossing
symmetry''.  In particular, consider a slice which would put two
vertices in the past and two in the future: by crossing symmetry,
H should allow not only (1,3) transitions, but also (2,2)
transitions.  (In a (2,2) transition the number of vertices does
not change, but the way
the vertices are coupled does change.)  

     A (2,2) slice is qualitatively different from a (1,3) slice,
however.  The (1,3) time slice cuts only the vertical, world line
edges of the tetrahedron; the (2,2) slice cuts both world line
edges and spin network edges.  In short, the tetrahedron is not
really 4-D symmetric: its sides are not all equivalent.  

     Markopoulou and Smolin \cite{MarkSmo} have proposed a
four-dimensional spin network formalism in which the fundamental
tetrahedrons are more fully symmetric, because all edges are
colored.  Markopoulou and Smolin do not use their formalism to
determine the form of the scalar constraint.  Indeed it would be
contrary to their philosophy to do so.  In their approach, the
classical Einstein-Hilbert action emerges at the {\it end} of a
long renormalization group calculation.  One starts from a
microscopic action (as yet unspecified, but presumably highly
symmetric).  If the initial action is in the right universality
class, then the renormalization group calculation yields correlated
behavior over macroscopic scales, as required by the classical
theory.  

     Sections I-III of this paper discusses three-dimensional spin
networks; but the results of those sections should carry over
readily to the 3+1 dimensional approaches just discussed. 
Presumably the initial microscopic
Markopoulou-Smolin action should be chosen so as to affect more
than one vertex at a time, insert no edges of color unity. and obey
a crossing symmetry (following Reisenberger and Rovelli) . 

     In order to obtain correlation and eliminate length
distortions, I have required that the color unity loop added by the
scalar constraint should be pushed outward from the vertex where
the constraint initially acts, until the loop fills an entire 
triangle of the spin network.   Consequently the constraint will
not change the number of vertices in the spin network (although it
can change the number of lines).  For example, the uncorrelated
scalar constraint action shown in figure~2 has added two new
vertices, whereas the correlated action shown in figure~5 adds no
new vertices (after recouplings which remove the color unity
lines).  

     To see how the constraint could add a new edge, set d = 0 in
figure~2 (no edge present initially). The two bottom vertices are
now divalent, but that can be remedied, if desired, by adding new
external lines; the valence will not matter.  As before, let the
constraint act at the top vertex, push the loop down the sides of
the triangle, and recouple so as to fill the entire triangle
(as in figures three through five, now with d = 0).  The d = 0 edge
is
replaced by $d'$ = 1; a new edge has been added.  

     One could demand that the procedure be modified so that not
even a new edge is added.  For example, suppose the triangle with
edges bcd in figure~2 were part of the larger spin network shown in
figure~8.  Again set d = 0 and let the scalar constraint act at the
upper vertex, but do not stop when the color
unity loop has filled the upper half triangle of figure~8; continue
to push this color unity loop until it fills the entire diamond.  
After recouplings such that $b \rta b', e \rta e',...$, there is 
neither a new vertex nor a new edge.      

     There is some rationale for stopping when the loop
has filled only the upper half of the diamond, however: if one
thinks of the spin
network as triangulating the space, then for pairs of edges which
share a common vertex, such as the pair (b,c) in figure~8, there
should be a third edge which connects the ends of b and c away from
the vertex, so as to form a triangle; otherwise the structure is
not rigid, in general.  One could say that the network of figure~8
with d = 0 really has a horizontal edge across the middle of the
diamond; the edge just happens to have the special color value
zero.  Put another way, the scalar constraint should leave the
number of
vertices fixed, while adding enough edges to form a rigid
structure.   This idea is appealing in its simplicity, but
sometimes an idea can be too simple for its own good.   Further
thought is needed.  

\begin{center}
{\bf Acknowledgements}
\end{center}

I would like to thank Lee Smolin for helpful discussions, and Lee
Smolin and Abhay Ashtekar for their hospitality while I was a
visitor at the Center for Gravitational Physics and Geometry.

\end{document}